\documentclass[aps,preprint,nofootinbib,superscriptaddress,prc]{revtex4}
\usepackage{amssymb}

\usepackage{epsfig}
\usepackage[bookmarksnumbered,bookmarksopen,colorlinks,citecolor=blue,linkcolor=blue]{hyperref}
\begin{document}

%\date{\today}

\title{Surface diffuseness correction in global mass formula}
\author{Ning Wang }
\thanks{wangning@gxnu.edu.cn}
\affiliation{Department of Physics, Guangxi Normal University,
Guilin 541004, P. R. China}

\author{Min Liu}
\affiliation{Department of Physics, Guangxi Normal University,
Guilin 541004, P. R. China}

\author{Xizhen Wu}
\affiliation{China Institute of Atomic Energy, Beijing 102413, P.
R. China}

\author{Jie Meng}
\thanks{mengj@pku.edu.cn}
\affiliation{State Key Laboratory of Nuclear Physics and Technology, School of Physics, Peking University, Beijing 100871, China}
\affiliation{School of Physics and Nuclear Energy Engineering, Beihang University, Beijing 100191, China}

\begin{abstract}
  By taking into account the surface diffuseness correction for unstable nuclei, the accuracy of the macroscopic-microscopic mass formula is further improved. The rms deviation with respect to essentially all the available mass data falls to 298 keV, crossing the 0.3 MeV accuracy threshold for the first time within the mean-field framework. Considering the surface effect of the symmetry potential which plays an important role in the evolution of the "neutron skin" toward the "neutron halo" of nuclei approaching the neutron drip line, we obtain an optimal value of the symmetry energy coefficient $J=30.16$ MeV. With an accuracy of 258 keV for all the available neutron separation energies and of 237 keV for the $\alpha$-decay Q-values of super-heavy nuclei, the proposed mass formula is particularly important not only for the reliable description of the \emph{r} process of nucleosynthesis but also for the study of the synthesis of super-heavy nuclei.
\end{abstract}

\maketitle

As one of the basic quantities in nuclear physics, the nuclear masses play key roles not only in the study of nuclear structure and reactions, but also in understanding the origin of elements in the universe. The nuclear mass formulas \cite{Lun03,Moll95,FRDM12,HFB17,Gor13,HFB27,Zhao10,Geng05,Meng13,Zhao1,GK,DZ28,Wang,Wang10,Liu11,Pom03} are of significant importance for describing the global nuclear properties and exploring the exotic structure of the extremely neutron-rich nuclei such as the halo phenomenon, the structure of super-heavy nuclei and their decay properties \cite{Ogan10,Naza,Sob,Zhou12}, and as well as the nuclear symmetry energy \cite{Dani14,LiBA,Tsang09,Liu10} which probes the isospin part of nuclear forces and intimately relates to the behavior of neutron stars. For finite nuclei, the diffuseness of the nuclear surface, which provides a measure of the thickness of the surface region and is intimately related to the nuclear surface energy \cite{Swi55}, is an important degree-of-freedom in the calculations of nuclear masses. The notations "neutron skin"  and "neutron halo" are adopted in Ref. \cite{Trz01} to describe the two extreme cases of two-parameter Fermi distributions of the neutron and proton peripheral density: the former refers to the case with equal diffuseness parameters for protons and neutrons and a larger half-density radius for the neutrons; the latter to the case with a much larger surface diffuseness for neutrons. For most stable nuclei, the corresponding density distribution is similar to the "neutron skin-type",  with a typical value around $0.5$ fm for the surface diffuseness. For nuclei near the neutron drip line, such as $^{11}$Li \cite{San06}, $^{22}$C \cite{Tanaka} and the giant-halo nuclei \cite{Meng98,Meng12}, the neutron matter extends much further, which implies the enhancement of the neutron surface diffuseness for these extremely neutron-rich nuclei. In nuclear mass calculations, all available global mass formulas, including the recent universal nuclear energy density functional (UNEDF) \cite{Erler12}, have not yet properly considered the surface diffuseness of exotic nuclei near the drip lines. It is well known that the symmetry energy plays an important role on the structure of neutron-rich nuclei. The thickness of neutron skin of nuclei has been explored to be linearly correlated with the slope of symmetry energy and the isospin asymmetry $I=(N-Z)/A$ of nuclei \cite{Trz01,Cent09}. On the other hand, the physics behind the skin and halo has been revealed as a spatial demonstration of shell effect from the relativistic continuum Hartree-Bogoliubov calculations \cite{Meng98a}. It is therefore necessary to investigate the influence of the surface diffuseness on the nuclear symmetry energy and shell correction for nuclei approaching the drip lines.

Inspired by the Skyrme energy-density functional, a macroscopic-microscopic mass formula, Weizs\"acker-Skyrme (WS) formula \cite{Wang,Wang10,Liu11}, was proposed with a rms deviation of 336 keV with respect to the 2149 measured masses \cite{Audi} in 2003 Atomic Mass Evaluation (AME). The Duflo-Zuker formula \cite{DZ28} with a rms deviation of 360 keV is also successful for the mass predictions. However, both of these two successful global mass formulas can not yet cross the 0.3 MeV accuracy threshold. In the WS formula, the axially deformed Woods-Saxon potential, as a phenomenological mean-field, is adopted to obtain the single-particle levels of nuclei. With the same value for the protons and neutrons, the surface diffuseness $a$ of the potential is set as a constant for all nuclei in the previous calculations. The obtained symmetry energy coefficient is about 29 MeV which is slightly smaller than the extracted one ($J\approx 30-32$ MeV) from some different approaches \cite{Dani14,LiBA,Tsang09,Gor13,Liu10,Stein,FRDM12,Latt12,Wang13}. The value of the symmetry coefficient can significantly affect the symmetry energy and thus the masses of nuclei near the neutron drip line. For example, the variation of the symmetry coefficient by one MeV can result in the variation of the symmetry energy by 33 MeV for the neutron-rich nuclei $^{176}$Sn. For more accurate description of the masses of drip line nuclei, it is required to further constrain the coefficient of the symmetry energy based on the new measured masses of nuclei far from stability. In this work, we attempt to further improve the WS formula by considering the nuclear surface diffuseness effect together with the latest nuclear mass datasets AME2012 \cite{Audi12}.

\begin{figure}
\includegraphics[angle=-0,width= 1.0\textwidth]{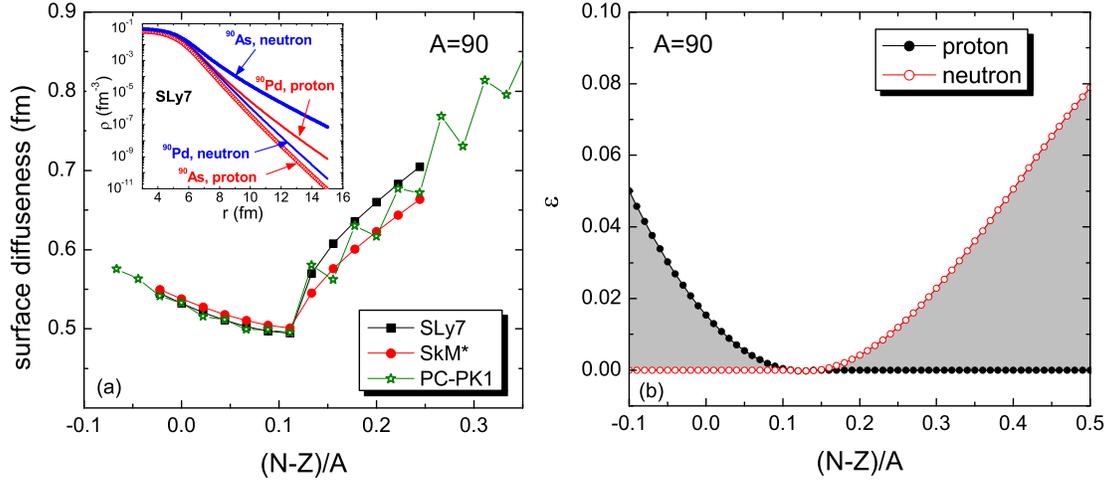}
 \caption{(Color online) (a) Calculated surface diffuseness of nuclei with $A=90$. The squares and circles denotes the results of the Skyrme energy density functional with SLy7 \cite{Chab1} and SkM* \cite{Bart82}, respectively. The stars denote the results of the relativistic density functional calculations with PC-PK1 \cite{Zhao10}, in which the staggering is due to the influence of the pairing in the PC-PK1 calculations. The inserted figure in (a) shows the density distributions of $^{90}$As and $^{90}$Pd with SLy7. (b) Correction factor $\varepsilon=(I-I_0)^2-I^4$ for the surface diffuseness in the single-particle potential with $I_0=0.4A/(A+200)$. The solid and open circles denote the results for protons and neutrons, respectively.   }
\end{figure}

To explore the correlation between the isospin asymmetry and the nuclear surface diffuseness, we first study the evolution of the nuclear density distribution for a series of isobaric nuclei by using the non-relativistic SLy7 \cite{Chab1} and SkM* \cite{Bart82} and the relativistic PC-PK1 \cite{Zhao10} density functionals.  From the mean-field point of view, the properties of all the nucleons in the nuclei are determined by the mean potential provided by their interaction with the other nucleons. Therefore the study of the isospin dependence of the potential, which become highly diffuse near the particle drip line, is crucial to understand unstable nuclei \cite{Meng99}. Fig. 1(a) shows the calculated nuclear surface diffuseness for nuclei with $A=90$ as a function of isospin asymmetry. Here, the value of nuclear surface diffuseness is extracted from fitting the calculated total density distribution $\rho(r)$ in the range of $r\le15$ fm with the Fermi function (under a logarithmic scale). Both the non-relativistic and relativistic density functional calculations show the enhancement of the nuclear surface diffuseness for nuclei far from stability. To illustrate this point more clearly, the sub-figure in Fig.1(a) shows the density distributions of $^{90}$As and $^{90}$Pd. For the neutron-rich nucleus $^{90}$As, the tail of the density distribution for the neutrons is much longer than that for the protons. For the proton-rich $^{90}$Pd, in contrast, the tail for the protons is just a little longer than that for the neutrons due to the Coulomb barrier. Simultaneously, we note that the surface diffuseness for protons (neutrons) in the neutron (proton)-rich nuclei do not change appreciately with the isospin asymmetry, which was also observed in the Sn and Pb isotopic chains \cite{Ward10}. The enhancement of the surface diffuseness for the very neutron-rich nuclei implies that the "neutron-skin" structure tends to evolve toward the "neutron-halo" structure for nuclei approaching the neutron drip line since the repulsion of the symmetry potential will "push" the extra-neutrons to the very low density region. At the neutron-deficient side, the extra-protons will be pushed to the surface region due to the Coulomb interaction and the symmetry potential.

Although the macroscopic-microscopic approaches are found to be the most accurate ones in the description of atomic masses \cite{Sob13}, the surface diffuseness effect for nuclei near the drip lines could affect the accuracy of the predictions. In the WS mass formula, the total energy of a nucleus is written as a sum of the liquid-drop energy, the Strutinsky shell correction
and the residual correction. The liquid-drop energy of a spherical nucleus $E_{\rm LD}(A,Z)$ is
described by a modified Bethe-Weizs\"acker mass formula,
\begin{eqnarray}
E_{\rm LD}(A,Z)=a_{v} A + a_{s} A^{2/3}+ E_C + a_{\rm sym} I^2 A f_{s} +
a_{\rm pair}  A^{-1/3}\delta_{np} + \Delta_W,
\end{eqnarray}
with the isospin asymmetry $I=(N-Z)/A$. $E_C=a_c \frac{Z^2}{A^{1/3}} \left ( 1- 0.76 Z^{-2/3} \right)$ and $\Delta_W$ denote the Coulomb energy term and the Wigner correction term for heavy nuclei \cite{Liu11}, respectively. The symmetry energy coefficient of finite nuclei is expressed as $a_{\rm sym}=c_{\rm sym} [1-\frac{\kappa}{A^{1/3}}+ \xi  \frac{2-|I|}{ 2+|I|A}  ]$ and the form of the correction factor $f_s$ for the symmetry energy will be presented in Eq.(6) and Fig. 1(b). The $a_{\rm pair}$ term empirically describes the odd-even staggering effect \cite{Wang}. Here, the $I^2$ term in the isospin dependence of $\delta_{np}$ is further introduced for a better description of the masses of even-A nuclei, with $\delta_{np}=|I|-I^2$ for the odd-odd nuclei and $\delta_{np}=(2-|I|-I^2) 17/16$ for the even-even nuclei.

To obtain the microscopic shell correction with the traditional Strutinsky procedure, the single particle levels of a nucleus
is calculated by using the code WSBETA \cite{Cwoik}.
The central potential $V$ is
described by an axially deformed Woods-Saxon form
\begin{eqnarray}
V (\vec {  r} \, )= \frac{V_q}{1+  \exp  [ \frac{  r -\mathcal {R}
(\theta)}{a}   ]}.
\end{eqnarray}
Where,  the depth $V_q$ of the central potential ($q=p$ for
protons and $q=n$ for neutrons) is written as
\begin{eqnarray}
 V_q = V_0 \pm V_s I
\end{eqnarray}
with the plus sign for neutrons and the minus sign for protons.
$V_s$ is the isospin-asymmetric part of the potential depth and has a value of $V_s=a_{\rm sym}$ in the WS formula, with which a unification of the microscopic and
the macroscopic parts is achieved. $\mathcal {R}$ defines the distance from the origin of
the coordinate system to the point on the nuclear surface
\begin{eqnarray}
\mathcal {R} (\theta)=c_0    R \, [1+  \beta_{2} Y_{20}(\theta) +
\beta_{4} Y_{40}(\theta) +\beta_{6} Y_{60}(\theta)],
\end{eqnarray}
with the scale factor $c_0$ which represents the effect of
incompressibility of nuclear matter in the nucleus and is
determined by the so-called constant volume condition
\cite{Cwoik}. $Y_{lm}(\theta, \phi)$ are the spherical harmonics.
$R=r_0 A^{1/3}$ denotes the radius of the single particle potential (with the same value for the protons and neutrons of a given nucleus). $a$ denotes the surface diffuseness of the potential. In this work, the isospin dependence of $a$ will be introduced and discussed later. For protons the Coulomb potential is additionally involved. The spin-orbit potential is written as
\begin{eqnarray}
V_{\rm s.o.}=-\lambda \left (\frac{\hbar}{2 M c} \right )^2\nabla
V \cdot (  \vec{ \sigma } \times  \vec{ p}),
\end{eqnarray}
where $\lambda= \frac{3}{2}\lambda_0 \left [ 1 \pm \frac{1}{3}( I - I^2) \right ]$  denotes the strength of the spin-orbit potential with the plus sign for neutrons and the minus sign for protons, in which the isospin dependence of the strength of the spin-orbit potential is considered according to the Skyrme energy-density functional. $M$ in Eq.(5) denotes the free nucleonic mass, $\vec{ \sigma }$ and $\vec{ p}$ are the Pauli spin matrix and the nucleon momentum, respectively.

The evolution of the "neutron-skin" toward the "neutron-halo" from the microscopic calculations indicates that the nuclear surface diffuseness is an important degree-of-freedom for the accurate descriptions of the ground state properties of nuclei near the drip lines. It is therefore necessary to introduce a surface diffuseness correction to the single-particle potential in the macroscopic-microscopic mass formula. In this work, the surface diffuseness $a$ of the Woods-Saxon potential in Eq.(2) is given by $a=a_0 \left (1+ 2 \varepsilon \delta_q \right ) $. Here, $\varepsilon=(I-I_0)^2-I^4$ denotes the correction factor [see Fig.1(b)] to the constant surface diffuseness $a_0$ of the Woods-Saxon potential. $I_0=0.4A/(A+200)$ denotes the isospin asymmetry of the nuclei along the $\beta$-stability line described by Green's formula. For nuclear matter, we assume $\varepsilon=0$ since the surface diffuseness disappears. $\delta_q=1$ for neutrons (protons) in the nuclei with $I>I_0$ ($I<I_0$), and $\delta_q=0$ for other cases. It means that the surface diffuseness of neutron distribution is larger than that of protons at the neutron-rich side and smaller than that of protons at the proton-rich side in the calculations. The shades in Fig.1(b) show the difference between the surface diffuseness of proton distribution and that of neutrons which strongly influences not only the nuclear symmetry energy in the macroscopic part, but also the shell correction in the microscopic part.

The corresponding correction $f_s$ for the symmetry energy and $f_d$ for the microscopic shell correction  due to the surface diffuseness are expressed as
\begin{eqnarray}
f_s=1+\kappa_s \varepsilon A^{1/3}
\end{eqnarray}
and
\begin{eqnarray}
 f_d=1+\kappa_d \varepsilon,
\end{eqnarray}
respectively. In the microscopic shell correction $\Delta E=c_1 f_d E_{\rm sh} + |I| E_{\rm sh}^{\prime}$ , $c_1$ is a scale factor \cite{Wang}, $E_{\rm sh}$ and $ E_{\rm sh}^{\prime}$ denote the shell energy of a nucleus and of its mirror nucleus obtained with the traditional Strutinsky procedure by setting the smoothing parameter $\gamma=1.2\hbar\omega_0$ and the order $p=6$ of the Gauss-Hermite polynomials. The $|I|$ term in $\Delta E$ is to take into account the mirror nuclei constrain \cite{Wang10} from
the isospin symmetry, with which the accuracy of the mass model can be improved by $10\%$. For stable nuclei, $f_s\simeq 1$ and $f_d\simeq 1$ according to Eqs.(6) and (7). The increase of the shades in Fig.1(b) represents the enhancement of nuclear symmetry energy for nuclei approaching the drip lines. We find that the surface diffuseness correction can significantly improve the accuracy of the predictions for the masses of the extremely neutron-rich and neutron-deficient nuclei.

\begin{table}
\caption{ Model parameters of the mass formula WS4. In addition to the model parameters mentioned in the text, $g_1$ and $g_2$ are the parameters related to the deformation energies of nuclei. The dependence of the macroscopic energy on the nuclear deformations in the WS formula is given by an analytical expression $E_{\rm LD} \prod  \left (1+b_k \beta_k^2 \right )$ with $b_k=\left ( \frac{k}{2} \right ) g_1A^{1/3}+\left ( \frac{k}{2} \right )^2 g_2 A^{-1/3}$ according to the Skyrme energy-density functional. $c_{\rm w} $ and $c_2$ denote the coefficient of the Wigner term and of the term for the residual mirror effect, respectively [see Eqs.(6) and (10) in Ref.\cite{Liu11} for details]. }
\begin{tabular}{cccc }
\hline\hline
  Parameter                         & ~~~~~~Value~~~~~~& ~~~ Parameter~~~       & ~~~Value~~~\\ \hline
 $a_v  \; $ (MeV)                  &   $-15.5181$   & $g_1 $                            &   0.01046 \\
 $a_s \; $  (MeV)                  &   17.4090      &  $g_2 $                           &   $-0.5069$ \\
 $a_c \; $ (MeV)                   &   0.7092       &  $V_0$ (MeV)                      &   $-45.8564$ \\
 $c_{\rm sym} $(MeV)               &   30.1594      &   $r_0$ (fm)                      &   1.3804   \\
 $\kappa \;  $                     &   1.5189       &   $a_0 $ (fm)                       &   0.7642   \\
 $\xi \;  $                        &   1.2230       &  $\lambda_0$                      &   26.4796  \\
  $a_{\rm pair} $(MeV)             &   $-5.8166$    &  $c_1  \; $                       &   0.6309   \\
  $c_{\rm w} $ (MeV)               &   0.8705       &  $c_2  \; ({\rm MeV} ^{-1})$      &   1.3371  \\
  $\kappa_{s} $                    &   0.1536       &  $\kappa_{d} $                    &   5.0086  \\
   \hline\hline
\end{tabular}
\end{table}

Based on the 2353 ($N$ and $Z\ge8$) measured nuclear masses $M_{\rm exp}$ in AME2012 \cite{Audi12} and searching for the minimal rms deviation with respect to the masses $ \sigma^2=  \frac{1}{m}\sum [M_{\rm exp}^{(i)}-M_{\rm th}^{(i)}]^2 $ by varying the values of the 18 independent model parameters, we obtain the optimal model parameters which is labelled as WS4 and listed in Table I. In the parameter searching procedure, the downhill searching method and the simulated annealing algorithm \cite{SA} are incorporated. The former is used for the parameters of the single-particle potential, while the latter is for the others. In Table II we list the rms deviations $\sigma (M)$ between the experimental masses and predictions of the models (in keV). The rms deviation with respect to essentially all the available mass data falls to 298 keV with the WS4 formula, the best value ever found within the mean-field framework. Comparing with the result of WS3, the value of $\sigma (M)$ is reduced by 37 keV. There are 219 "new" data for nuclei first appearing in AME2012 [see the solid squares in Fig.2(b)] and generally far from the $\beta$-stability line. Considering the surface diffuseness effect, the rms deviation with respect to the masses of these 219 nuclei falls to 346 keV. Comparing with the result of WS3, the improvement is 78 keV, which is significantly larger than the average improvement of 37 keV for the $\sigma (M)$. Similarly, the masses for the 286 nuclei with $|I-I_0|>0.1$ [see the crosses in Fig.2(b)] are much better reproduced with the WS4 formula. In addition, as one of the most prominent global interpolation and extrapolation schemes, the radial basis function (RBF) approach is powerful and efficient for further improving the accuracy of the global nuclear mass formulas \cite{Wang11,Niu13}. Based on the WS4 calculations together with the RBF corrections proposed in \cite{Wang11}, the rms deviation with respect to all the 2353 masses remarkably falls to 170 keV and the rms deviation to the 219 "new" data falls to 155 keV, approaching the chaos-related unpredictability limit ($\sim 100 $ keV) for the calculation of nuclear masses \cite{Barea05}.

\begin{table}
\caption{ Rms deviations between data and
predictions from the WS4 formula (in keV). The line $\sigma(M)$ refers to all the 2353 measured masses in AME2012, the
line $\sigma (M_{\rm new})$ to the measured masses of 219 "new" nuclei in AME2012, the line $\sigma (M_{0.1})$ to the masses of 286 nuclei with $|I-I_0|>0.1$, the line $\sigma (S_n)$ to all the 2199 measured neutron separation energies $S_n$, the line $\sigma (Q_\alpha)$ to the $\alpha$-decay energies of 46 super-heavy nuclei ($Z\ge 106$) \cite{Wang10}. The corresponding results of WS3 model are also presented for comparison.  WS4$^{\rm \bf{RBF}}$ denotes that the radial basis function (RBF) corrections \cite{Wang11} are combined in the WS4 calculations. }
\begin{tabular}{cccc}
 \hline\hline
                          & WS3  & ~~WS4~~ & WS4$^{\rm \bf{RBF}}$ \\
\hline
 $\sigma  (M)$           & $335$ & $298$ &  $170 $\\
 $\sigma  (M_{\rm new})$ & $424$ & $346$ &  $155 $\\
 $\sigma  (M_{\rm 0.1})$  & $516$ & $444$ & $215 $\\
 $\sigma  (S_n)$         & $273$ & $258$ &  $251 $\\
 $\sigma  (Q_\alpha)$    & $248$ & $238$ &  $237 $\\
\hline\hline
\end{tabular}
\end{table}

\begin{figure}
\includegraphics[angle=-0,width=1.0\textwidth]{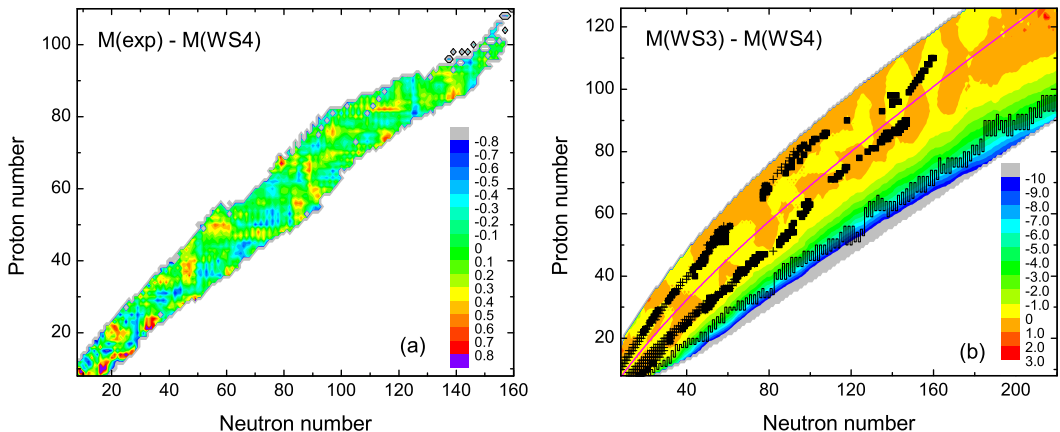}
 \caption{(Color online) (a) Difference between measured and calculated masses with the WS4 formula. (b) Difference between calculated masses with WS3 and those with WS4. The squares and crosses denote the nuclei first appearing in AME2012 and the nuclei with $|I-I_0|>0.1$, respectively. The smooth and the zigzag curve denote the $\beta$-stability line from Green's formula and the neutron drip line from the WS4 formula, respectively. }
\end{figure}

In Fig.2(a), we show the deviations of the calculated masses from the experimental data. For all the 2353 nuclei with $Z$ and $N\ge 8$, no deviation exceeds 1.23 MeV. Fig.2(b) shows the difference between the calculated masses with WS3 and those with WS4. For most nuclei, the results of WS3 and WS4 are consistent in general (with  deviations smaller than one MeV). For nuclei near the neutron drip line, the masses given by WS4 are larger than the results of WS3 by several MeV. This is due to the enhancement of the nuclear symmetry energy coming from the surface diffuseness effect in the extremely neutron-rich nuclei. With the correction $f_s$ for the symmetry energy term, we also note that the bulk symmetry energy coefficient $c_{\rm sym}$ increases by one MeV, up to 30.16 MeV. The value of $c_{\rm sym}$  represents the symmetry coefficient $J$ of nuclear matter at saturation density. The optimal value of $J=30.16$ MeV for the symmetry energy is consistent with the value of 30.0 MeV suggested in the latest Skyrme Hartree-Fock-Bogoliubov (HFB) mass formulas \cite{HFB27,Gor13} in which the model parameters are determined from the same nuclear mass datasets.

In summary, the surface diffuseness effect of nuclei near the drip lines is taken into account for the first time in the macroscopic-microscopic mass calculations. The rms deviation with respect to the 2353 known masses falls to 298 keV, the best value ever found within the mean-field approximation. The surface diffuseness of drip line nuclei influences both the symmetry energy and the shell corrections. With the surface diffuseness correction for unstable nuclei, we obtain an optimal value of 30.16 MeV for the symmetry energy coefficient which is consistent with the value in the latest Skyrme HFB formulas. The systematic improvement for the masses of neutron-rich nuclei demonstrates the recoupling of the proton and neutron matter and implies the possible existence of the "neutron halo" structure for nuclei approaching the neutron-drip line, especially for the light nuclei in which the value of the diffuseness correction $\varepsilon$ is quite large due to the extremely large value of $N/Z$.

\begin{center}
\textbf{ACKNOWLEDGEMENTS}
\end{center}

We thank Zhu-Xia Li and an anonymous referee for valuable comments and suggestions, and Ying Chen for providing the results of the relativistic mean-field calculations. This work was supported by National Natural Science Foundation of China (Nos 11275052, 11365005, 11335002, 91126010, 11375062), Defense Industrial Technology Development Program (No. B0120110034) and National Key Basic Research Program of China (Grant No. 2013CB834400). The nuclear mass tables with WS formula are available from http://www.imqmd.com/mass/


\begin{thebibliography}{99}


\bibitem{Lun03} D. Lunney, J.M. Pearson, C. Thibault, Rev. Mod. Phys. \textbf{75}, 1021 (2003).

\bibitem{Moll95} P. M\"oller, J. R. Nix, W. D. Myers, W. J. Swiatecki, At. Data and
Nucl. Data Tables \textbf{59}, 185 (1995).

\bibitem{FRDM12}  P. M\"oller, W. D. Myers, H. Sagawa, and S. Yoshida, Phys. Rev.
Lett. \textbf{108}, 052501 (2012).

\bibitem{HFB17} S. Goriely, N. Chamel and J. M. Pearson, Phys. Rev. Lett. \textbf{102}, 152503 (2009).

\bibitem{Gor13} S. Goriely, N. Chamel, and J. M. Pearson, Phys. Rev. C \textbf{88}, 024308 (2013).

\bibitem{HFB27} S. Goriely,  N. Chamel,  and J. M. Pearson, Phys. Rev. C \textbf{88}, 061302(R) (2013)

\bibitem{Zhao10} P. W. Zhao, Z. P. Li, J. M. Yao, and J. Meng, Phys. Rev. C \textbf{82},  054319 (2010).

\bibitem{Geng05} L. S. Geng, H. Toki, J. Meng, Prog. Theor. Phys. 113, 785 (2005).

\bibitem{Meng13} J. Meng, J. Peng, S. Q. Zhang, et al., Front. Phys. \textbf{8}, 55 (2013).

\bibitem{Zhao1} H. Jiang, G. J. Fu, Y. M. Zhao, and A. Arima, Phys. Rev. C \textbf{82}, 054317 (2010).

\bibitem{GK} J. Barea,  A. Frank,  J. G. Hirsch, P. Van Isacker, S. Pittel, and V. Vel\'azquez,  Phys. Rev. C \textbf{77},
041304 (2008).

\bibitem{DZ28} J. Duflo and A. P. Zuker, Phys. Rev. C \textbf{52}, 23 (1995).

\bibitem{Wang} N. Wang, M. Liu and X. Z. Wu, Phys. Rev. C \textbf{81}, 044322 (2010).

\bibitem{Wang10} N. Wang, Z. Y. Liang, M. Liu and X. Z. Wu, Phys. Rev. C \textbf{82}, 044304 (2010).

\bibitem{Liu11} M. Liu, N. Wang, Y. G. Deng, and X. Z. Wu, Phys. Rev. C \textbf{84}, 014333 (2011).

\bibitem{Pom03} K. Pomorski and J. Dudek, Phys. Rev. C \textbf{67}, 044316 (2003).

\bibitem{Ogan10} Yu. Ts. Oganessian et al.,  Phys. Rev. Lett.
\textbf{104}, 142502 (2010).

\bibitem{Naza} S. Cwiok, P. H. Heenen and W. Nazarewicz, Nature \textbf{433}, 705 (2005).

\bibitem{Sob} A. Sobiczewski,  K. Pomorski, Prog. Part. Nucl. Phys. \textbf{58},
292 (2007).

\bibitem{Zhou12} B. N. Lu, E. G. Zhao, and S. G. Zhou, Phys. Rev. C 85, 011301(R) (2012).


\bibitem{Dani14} P. Danielewicz and J. Lee, Nucl. Phys. A {\bf 922}, 1 (2014).

\bibitem{LiBA} B. A. Li, L. W. Chen, C. M. Ko, Phys. Rep. \textbf{464}, 113
(2008).

\bibitem{Tsang09} M. B. Tsang, Y. X. Zhang, P. Danielewicz, M. Famiano, Z. Li,
W. G. Lynch, and A. W. Steiner,   Phys. Rev. Lett.  {\bf 102},
122701 (2009).

\bibitem{Liu10} M. Liu, N. Wang, Z. X. Li, and F. S. Zhang, Phys. Rev. C \textbf{82}, 064306 (2010).


\bibitem{Swi55} W. J. Swiatecki, Phys. Rev. \textbf{98}, 203 (1955).

\bibitem{Trz01} A. Trzci\'nska, J. Jastrzebski, and P. Lubi\'nski, F. J. Hartmann, R. Schmidt, T. von Egidy, B. K\l{}os, Phys. Rev. Lett. \textbf{87}, 082501 (2001).

\bibitem{San06} R. S\'anchez, W. N\"ortersh\"auser, G. Ewald, D. Albers, J. Behr, P. Bricault, B. A. Bushaw, A. Dax, J. Dilling, et al., Phys. Rev. Lett. \textbf{96}, 033002 (2006).

\bibitem{Tanaka} K. Tanaka, T. Yamaguchi et al., Phys. Rev. Lett. \textbf{104}, 062701 (2010).

\bibitem{Meng98} J. Meng and P. Ring, Phys. Rev. Lett. 80, 460 (1998).

\bibitem{Meng12} Y. Zhang, M. Matsuo, and J. Meng, Phys. Rev. C \textbf{86}, 054318 (2012).

\bibitem{Erler12} J. Erler, N. Birge, M. Kortelainen, W. Nazarewicz, E. Olsen, A. M. Perhac, and M. Stoitsov, Nature \textbf{486}, 509 (2012).

\bibitem{Cent09} M. Centelles, X. Roca-Maza,  X. Vin\~as,  and M. Warda, Phys. Rev. Lett. \textbf{102}, 122502 (2009).

\bibitem{Meng98a} J. Meng, I. Tanihata and S. Yamaji, Phys. Lett. B \textbf{419}, 1 (1998).


\bibitem{Audi} G. Audi, A.H. Wapstra and C. Thibault, Nucl. Phys. A \textbf{729}, 337 (2003).

\bibitem{Stein} A. W. Steiner and S. Gandolfi, Phys. Rev. Lett. \textbf{108}, 081102 (2012).

\bibitem{Latt12} J. M. Lattimer, Annu. Rev. Nucl. Part. Sci. \textbf{62}, 485 (2012).

\bibitem{Wang13} N. Wang, L. Ou, and M. Liu, Phys. Rev. C \textbf{87}, 034327 (2013).

\bibitem{Audi12} G. Audi, M. Wang, A. H. Wapstra, F. G. Kondev, M. Mac-Cormick, X. Xu, and B. Pfeiffer, Chin. Phys. C \textbf{36}, 1287 (2012).

\bibitem{Chab1} E. Chabanat, P. Bonche, P. Haensel, J. Meyer, R. Schaeffer, Nucl. Phys. A \textbf{627}, 710 (1997).

\bibitem{Bart82} J. Bartel, P. Quentin, M. Brack, C. Guet, H.-B. Hakansson, Nucl. Phys. A  \textbf{386}, 79 (1982).

\bibitem{Meng99} J. Meng, I. Tanihata, Nuclear Physics A \textbf{650}, 176 (1999).

\bibitem{Ward10} M. Warda,  X. Vin\~as,  X. Roca-Maza,  and M. Centelles, Phys. Rev. C \textbf{81}, 054309 (2010).

\bibitem{Sob13} A. Sobiczewski  and Yu. A. Litvinov, Phys. Scr. \textbf{T154}, 014001 (2013); Phys. Rev. C \textbf{89}, 024311 (2014).

\bibitem{Cwoik} S. Cwoik, J. Dudek, W. Nazarewicz, J. Skalski, and T. Werner, Comp. Phys. Comm. \textbf{46}, 379 (1987).


\bibitem{SA} A. Corana, M. Marchesi, et al., ACM Transactions on Mathematical
Software, \textbf{13}, 262 (1987).

\bibitem{Wang11} N. Wang and M. Liu, Phys. Rev. C \textbf{84}, 051303(R)
(2011); http://www.imqmd.com/mass/

\bibitem{Niu13} Z. M. Niu, Z. L. Zhu, Y. F. Niu, B. H. Sun, T. H. Heng, and J. Y. Guo, Phys. Rev. C \textbf{88}, 024325 (2013)

\bibitem{Barea05} J. Barea, A. Frank, and J. G. Hirsch, Phys. Rev. Lett. \textbf{94}, 102501 (2005)



\end{thebibliography}
\end{document}